\documentclass{ar-1col}

\usepackage[comma]{natbib}
\usepackage{url}

\usepackage{amsmath}
\usepackage{amssymb}
\usepackage{bm}
\usepackage{bbm}
\usepackage{enumitem}

\DeclareMathOperator{\E}{\mathbb{E}}
\DeclareMathOperator{\p}{\mathbb{P}}
\DeclareMathOperator{\indep}{\perp\!\!\!\perp}

\DeclareMathOperator*{\argmin}{arg\,min}

\jname{Xxxx. Xxx. Xxx. Xxx.}
\jvol{AA}
\jyear{YYYY}
\doi{10.1146/((please add article doi))}

\begin{document}

\markboth{Slavkovi\'c and Seeman}{Statistical Data Privacy}

\title{Statistical Data Privacy: A Song of Privacy and Utility}

\author{Aleksandra Slavkovi\'c$^1$ and Jeremy Seeman$^2$
\affil{$^1$Department of Statistics, The Pennsylvania State University, University Park, PA, 16802; email: abs12@psu.edu}
\affil{$^2$Department of Statistics, The Pennsylvania State University, University Park, PA, 16802, email: jhs5496@psu.edu. ORCID: 0000-0003-3526-3209}}

\begin{abstract}

To quantify trade-offs between increasing demand for open data sharing and concerns about sensitive information disclosure, statistical data privacy (SDP) methodology analyzes data release mechanisms which sanitize outputs based on confidential data. Two dominant frameworks exist: statistical disclosure control (SDC), and more recent, differential privacy (DP). Despite framing differences, both SDC and DP share the same statistical problems at its core. For inference problems, we may either design optimal release mechanisms and associated estimators that satisfy bounds on disclosure risk, or we may adjust existing sanitized output to create new optimal estimators. Both problems rely on uncertainty quantification in evaluating risk and utility. In this review, we discuss the statistical foundations common to both SDC and DP, highlight major developments in SDP, and present exciting open research problems in private inference.

\end{abstract}


\maketitle

\tableofcontents

\section{INTRODUCTION}

\subsection{Statistical data privacy}

Privacy and confidentiality (which we, in this review, synonymously refer to as ``privacy") are widely viewed as an essential component of free societies \citep{westin1968privacy,cohen2012privacy}. As large-scale data collection becomes more commonplace, threats to individual's data privacy grow ever more prominent (e.g., see  \cite{dwork2017exposed} for a survey of common attacks). Despite these threats, many statisticians and data users, have a limited understanding of what data privacy is and how it affects our work. Data privacy is often synonymous with boilerplate procedures required to satisfy compliance obligations (e.g., HIPPA, IRBs), an inconvenience to our normal operating procedures. We may think that the solution is always anonymization, or the act of removing personally identifying information (PII) from a database. Yet statistical outputs (such as summary statistics and parameter estimates) pose threats to individual disclosure, just as their inputs do through access to confidential databases; this makes data privacy a methodological problem beyond the known failures of anonymization alone \citep{ohm2009broken}. Not all statistics reveal the same information about individuals, and negotiating to find the proper balance between privacy and utility (via uncertainty quantification) is precisely where statistics has much to offer.

Statistical data privacy (SDP) aims to develop provable and usable data privacy theory and methodology, by integrating tools from computer science and statistics, to enable broad sharing of data across many different data context and domains where it is desired or required that individual’s identities or sensitive attributes are protected; e.g., census, health, genomic data, social networks. SDP methods need to minimize privacy loss/disclosure risk of sensitive information while at the same time preserve sufficient statistical integrity of data in order to support valid inference (i.e., maximize data utility).  
Two dominant frameworks in SDP, defined by different units of analysis, make different conceptual trade-offs about adversarial assumptions, disclosure risks that is privacy definitions, and their effects on downstream inference. Statistical disclosure control (SDC) or limitation (SDL) methods (e.g., \cite{hundepool2012statistical}) typically analyze individual databases, whereas differential privacy (DP) methods \citep{dwork2014algorithmic} analyze pairs of databases in a shared schema (space of possible databases). 

Despite good-faith attempts to unify these perspectives, some dating back to the onset of DP, as discussed in \cite{slavkovic2013}, SDC and DP research perspectives still diverge. SDP scholars, data administrators, and quantitative social scientists, it seems, have become more polarized as proponents of one perspective or the other. In particular, the U.S. Census Bureau's decision to use DP has led to debates about the merits of either approach \citep{abowd2021third,domingo2021limits}. Furthermore, this insularity has lead to growing gaps between theoretical developments, now dominated by DP research, and applied methodology, now dominated by SDC research. Such debates remain important, as they highlight the inherently political nature of privacy-preserving data stewardship \citep{rogaway2015moral,boyd2022differential} and the way mathematical formalisms frame discussions about privacy \citep{seeman2022politics}. Still, SDC and DP share more similarities than differences, and unifying these ideas can help mutually enrich future SDP research.

In this article, we have three main goals: first, we highlight how SDC and DP methods are built upon common statistical foundations that make different but necessary compromises in conceptualizing privacy. Second, we discuss how SDP is inseparable from the study of data generating processes in both designing optimal private estimators and adjusting inferences for privacy preservation given sanitized outputs. Third, we showcase open statistical problems typically left unarticulated by theoretical SDP research, such as valid statistical inference, computational tractability, and compatibility with probability models. 
The paper is organized as follows: Sections \ref{sec:sdc} and \ref{sec:fp} review privacy quantification and release mechanisms/methods using SDC and DP, respectively. Section \ref{sec:sdp_shared} reviews inferential problems common to all SDP. Section \ref{sec:case_study} presents a fabulistic case study applying SDP methodology to a small population of the Westerosis. Finally, Section \ref{sec:disc} discusses current research practices and proposes future research directions to improve SDP research.

\subsection{Notation and Problem Formulation}
\label{sec:prob_form}

Throughout this paper, we use the following notation and terminology. Let $\mathcal{X}$ be the sample space for one individual's contributions to a database, and let $\mathcal{D} = \mathcal{X}^n$ be the sample space for a confidential database of $n$ individuals who have contributed their data. We refer to $\mathcal{D}$ as the schema, or the space of possible databases. For the purposes of this review, we assume the unit of observation refers to one individual; however, similar methodology may be applied to groups, businesses, or organizations.

\begin{marginnote}
\entry{Schema}{the space of all possible confidential databases, $\mathcal{D}$.}
\entry{Output space}{$\mathcal{S}$, the space of all possible statistical outputs.}
\entry{Release mechanism (RM)}{a randomized function $S$ from $\mathcal{D}$ to $\mathcal{S}$ which sanitizes a database $D$'s output.}
\entry{Sanitized output}{$S(D)$, a statistical output from a confidential database that has undergone some form of privacy preservation.}
\end{marginnote}

In SDP, we release a statistic from a set of possible statistical outputs, which we call the output space, $\mathcal{S}$. A release mechanism (RM), defined by $S: \mathcal{D} \mapsto \mathcal{S}$, is a transformation of the confidential data which produces the sanitized output, $S(D)$. SDP characterizes the privacy risk and data utility properties of different RMs in different scenarios. RMs may be either deterministic, i.e., $S(D)$ transforms the confidential database $D$ according to a fixed function such as aggregation, or randomized, i.e., $S(D)$ is a random variable that varies conditionally on $D$ such as a statistic with randomized noise. Moreover, $\mathcal{S}$ can take many forms, ranging from synthetic or tabular microdata releases to summary statistics and model parameter estimates. SDC and DP typically ask two different questions about the relationship between the RM $S$ and the database $D \in \mathcal{D}$ with respect to disclosure risk: 
\begin{marginnote}
\entry{Microdata}{Record-level data, either directly from the confidential database, or synthetically generated from a model.}
\end{marginnote}
\begin{enumerate}
    \item \textbf{SDC perspective}: how does a particular statistical release $S(D)$ for a particular database $D \in \mathcal{D}$ limit a particular measure of disclosure risk, with respect to an individual or sub-population, dependent on existing adversary knowledge?
    \item \textbf{DP perspective}: how can a statistical RM $S$ limit the ability for adversaries to distinguish between similar databases $D, D' \in \mathcal{D}$, such as by only changing one individual's contribution, within the same schema $\mathcal{D}$?
\end{enumerate}
In asking these questions, SDC and DP frame the problem of data privacy with different conceptual compromises. SDC defines a narrow set of adversarial contexts with the goal of quantifying a risk measure on a particular database, which we refer to as an absolute measure of disclosure risk. Alternatively, DP defines a broader set of adversarial contexts with the goal of quantifying a measure of disclosure risk differences for any two similar databases within the schema, which we refer to as a relative measure of disclosure risk.

SDC and DP are both on a spectrum of many possible ways to reason about SDP, but both require negotiating between privacy and utility. As the number of statistics released about any database increases, one can increasingly reconstruct individual records contained in that database \citep{dinur2003revealing}, (this problem, in part, motivated the start of DP research). At the same time, assumptions about data generation processes and context-specificity are necessary to provide meaningful data utility for any such statistical data release (\cite{kifer2011no}, A.K.A. "no free lunch in data privacy). Therefore it behooves us to investigate both SDC and DP simultaneously.

For historical context, SDC methods have been used in official statistics since the 1960s \citep{mckenna2019disclosure}. Early attempts to mathematically formalize SDP date back to the seminal works of Tore Dalenius \citep{dalenius1977towards}, who viewed privacy as striving to reduce the harms of population-level statistical inferences on individuals:

\begin{extract}
“If the release of the statistics $T(D)$ makes it possible to
determine the value [of confidential statistical data] more accurately than is possible without access to $T(D)$, a disclosure has taken place.”
\end{extract}

However, modern literature in both SDC and DP has shown that we cannot learn anything about a population without also learning something about individuals within that population \citep{dwork2010difficulties}, and thus, the above definition is unobtainable. As an oft-cited counterexample from \cite{dwork2017exposed},

\begin{extract}
``Releasing the fact that smoking and lung cancer are strongly correlated reveals sensitive
information about any individual known to smoke; however, we do not consider this to be a privacy violation, as learning this correlation has nothing to do with the use of that individual’s data."
\end{extract} 
Such concerns have become a point of confusion due to misinterpretations of DP \citep{kenny2021impact}, leading some in DP to recently restate this commitment \citep{ullman2021statistical}. Here, we further clarify that such a feature is central to the entire project of SDP, not solely DP. 

\begin{figure}
    \centering
    \includegraphics[width=.90\textwidth]{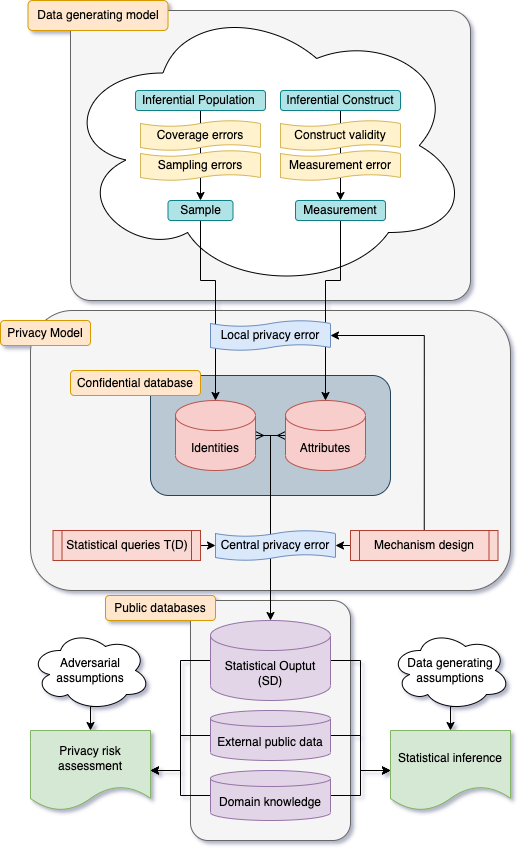}
    \caption{Flowchart of SDP modeling for privacy risk and data utility assessment}
    \label{fig:sdp_flowchart}
\end{figure}

So, how does SDP becomes part of the broader project of statistical inference? Figure \ref{fig:sdp_flowchart} graphically illustrates where privacy preservation sits in statistical inference, with a special focus on the social science context. Because we are working with human data, we are prone to many error sources, as often systematized by survey methodology \citep{Groves2011}. These errors shape the structure of our confidential database records, even before any sanitization is introduced. 

Note that we have two kinds of errors due to privacy, which enter the data generating process at different stages and correspond to different database trust models \citep{stoller2011trust}. In the local model, privacy protecting errors are introduced into the way users contribute to the confidential database \citep{evfimievski2003limiting}. Alternatively, in the central model \citep{dwork2014algorithmic}, user contributions are combined and transformed into sanitized releases. Local models confer stronger privacy guarantees, because unlike the central model, certain information about users is inaccessible even to the data curator. 

We argue that the statistical perspective is essential because both risk assessment and inference on statistical output relies on the probabilistic transformations throughout the data generating process that influence said statistical output. Much of the focus on SDP research narrowly considers only the privacy model, i.e., only the relationships between the confidential database and the statistical outputs. By taking a bird's eye view of this process in Figure \ref{fig:sdp_flowchart}, we see many possible avenues for statistically motivated privacy research. These research avenues depend on where we as statisticians are involved in the process. Namely, do we have any say in choosing RM? Depending on the answer, we can consider two different broad classes of problems. 

\begin{enumerate}
    \item \textbf{Design problem}: if $\mathcal{Q}$ is a class of RMs with the same privacy guarantees, how do we find an optimal RM $S^* \in \mathcal{Q}$, associated optimal estimator $\hat{\theta}^*(S^*(D))$ for some $\theta \in \Theta$, and what is the uncertainty in $\hat{\theta}^*$?
    \item \textbf{Adjustment problem}: given a sanitized statistical result $S(D)$, how do we find an optimal estimator $\hat{\theta}^*(S(D))$ for some $\theta \in \Theta$, and what is the uncertainty in $\hat{\theta}^*$?
\end{enumerate}

Both approaches will require different means of uncertainty quantification, as we have more flexibility when we get to decide the form of $S$. Still, both problem classes are equally important, since we as statisticians may be working directly with confidential data, or we may be working with private synthetic data.

\section{Statistical disclosure control (SDC)}
\label{sec:sdc}

SDC operationalizes the trade-off between risk and utility within the context of a single observed database $D \in \mathcal{D}$. Data curators construct RMs $S(D)$, and analyze their privacy properties using a disclosure risk measure (DRM), $R: \mathcal{S} \mapsto \mathcal{R}$. The RM $S$ is then altered depending on both the data utility offered by $S(D)$ compared to $D$, and the risk $R(S(D))$ compared to $R(D)$. In doing so, the confidential data is reused by the data curator multiple times in order to calibrate the balance between privacy and data utility. 
\begin{marginnote}
\entry{Disclosure risk measure (DRM)}{a function $R$ from $\mathcal{S}$ to $\mathcal{R}$ which quantifies the disclosure risk of a sanitized output, $R(S(D))$}
\end{marginnote}

\subsection{SDC methods}

As originally formulated by \cite{duncan1991enhancing}, SDC methods for matrix-valued databases often belong to a class of linear transformations, i.e., $S(D) \triangleq ADB + C.$ Here, $A$ describes a record-level transformation, $B$ describes variable-level transformations, and $C$ describes additional displacement or randomized noise. Note that in practice, these transformations need not be linear, nor randomized. Regardless, each of these transformation classes introduces baise and variance to the confidential data, and thus new considerations into the data analysis process, which we briefly review:

\begin{itemize}
    \item \textbf{Record-level transformations} include members of confidential database in the released statistics with varying probabilities. Common approaches involve random sampling, outlier removal, and special unique removal \citep{willenborg1996statistical}. 
    \item \textbf{Variable transformations} typically shrink $\mathcal{S}$ relative to $\mathcal{D}$.  For tabular data, cell suppression, recoding, top-coding, and aggregation are all examples of schema reduction techniques that reduce $\mathcal{S}$ \citep{hundepool2012statistical}. 
    \item \textbf{Randomized masking} injects randomized noise into quantitative statistics to prevent direct inference on any statistic exactly calculated from individual records. Note that even though randomization is a central component of DP, randomized SDC methods date back to the 1960s with randomized response \citep{warner1965randomized}.
\end{itemize}

As a generalization of Duncan et al's approach, synthetic data generation methods produce sanitized output similar to $D$ but with records randomly sampled. This sampling can occur for part of any individual record or for the entire record, and we could produce partially or fully synthetic data. For more details, see \cite{drechsler2010sampling}, \cite{snoke2018general}, and a recent Annual Review article by \cite{raghunathan2021synthetic}.

\subsection{Disclosure risk measures (DRMs)}
\label{sdc:risk}
Given these methods, we now turn to what $R(S(D))$ means in practice. The DRM $R$ captures information about individuals that can be inferred from the statistical release $S(D)$. Choices for $R$ make different implicit and explicit assumptions about what adversaries know in advance, what constitutes statistical disclosure, and how to quantify the probability of those potential disclosures. In general, there are three broad categories of DRMs: 

\begin{enumerate}
    \item \textbf{Quasi-identifiability measures}: Quasi-identifiability is the ability for combinations of certain covariates to isolate individuals in the dataset. As an example, $S$ satisfies $k$-anonymity \citep{sweeney2002k} if:
    \begin{equation}
        \label{eq:k_anon}
        R(S(D)) = \min_{X \in \mathcal{X}} \#\{ i \in [n] \mid X_i = X \} \geq k.
    \end{equation}
    Notable variants include $\ell$-diversity \citep{machanavajjhala2007diversity} and $t$-closeness \citep{li2007t}, which extend $k$-anonymity to capture the heterogeneity of sensitive user contributions within quasi-identifying categories. For databases with non-discrete entries, alternative approaches may be used based on clustering, microaggregation, or outlier detection \citep{domingo2002practical}. 
    
    \item \textbf{Model-based reidentification measures:} SDC methods often involve modeling whether particular entries are reidentifiable under various modeling assumptions in the worst-case scenario (but still within the context of a single database, unlike DP). We define this event as $r_i = 1$ for $i \in [n]$ based on a probability model $\p_\theta$ for $\theta \in \Theta$. This allows us to construct reidentification rates of the form:
    \begin{equation}
        \label{eq:sdc_reid_risk}
        R(S(D)) = \frac{1}{n} \sum_{i=1}^n \sup_{\theta \in \Theta} \p_\theta( r_i = 1 \mid S(D) ).
    \end{equation}
    The effectiveness of the measure depends on the model accuracy for $\p(r_i = 1 \mid S(D))$. When $\mathcal{X} = k$, different techniques can be used to model the joint distribution of frequencies for categories in the population and sample \citep{franconi2004individual}; e.g., these can be based on log-linear models \citep{fienberg1998disclosure,skinner2008} or survey estimation techniques \citep{skinner2009statistical}. The formula in Equation \ref{eq:sdc_reid_risk} calculates an average reidentification rate; however we may interested in other summary statistics of the individual reidentification probabilities $\p(r_i = 1 \mid S(D))$, such as their maximum in a worst-case analysis.
    
    \item \textbf{Data-based reidentification measures:} while theoretical models can upper bound DRMs, we can alternatively lower bound these risks by attempting such database reconstruction attacks with external data sources \citep{domingo2003disclosure,winkler2004}. The DRM is then a linkage rate, or:
    \begin{equation}
        R(S(D)) = \frac{1}{n} \sum_{i=1}^n \mathbbm{1}\{\text{Record $X_i \in D$ successfully linked to a record $Z_j \in Z$} \}.
    \end{equation}
    Such an approach depends on multiple factors: how are potential records $X_i \in D$ extracted from $S(D)$? What determines a successfully linked record? And how does the external data $Z$ relate to the population? Such questions are answered by fundamental connections between database reconstruction and record linkage \citep{Dobra2009,vatsalan2013taxonomy,garfinkel2019understanding}.
\end{enumerate}

In the current data landscape, however, there are systemic downsides to using SDC that could be viewed as weaknesses of the framework. First, data curators often cannot disclose the mathematical form of the RM $S$ without leaking additional confidential information (e.g., see \cite{drechsler2010sampling} on data swapping, or \cite{slavkovic2004thesis} for cell suppression, which also shows how lack of transparency negatively impacts statistical inference). Next, SDC methods are not robust to post-processing: there could be transformations of our releases $h$ where $h(S(D))$ and $S(D)$ have different DRMs. Finally, SDC methods do not easily compose, in that if we have two release strategies $S_1(D)$ and  $S_2(D)$ and we know their risks, it may be difficult to quantify the risk of the joint release $(S_1(D), S_2(D))$. 

\section{Differential privacy (DP)}
\label{sec:fp}
Differential privacy (DP) is a framework which mathematically formalizes the privacy properties of data release strategies and addresses the above shortcomings. By starting with a privacy definition and necessitating additional randomness, DP methods are provably consistent with the privacy definition and able to satisfy these three properties. 
\begin{enumerate}
    \item \textbf{Methodologically transparent:} knowledge of $S$ preserves $S(D)$'s privacy risk. 
    \item \textbf{Robust to post-processing:} $h(S(D))$ has, at most, the same privacy risk as $S(D)$.
    \item \textbf{Composable:} we can analytically express the privacy risk of two DP releases $S_1(D)$ and $S_2(D)$ when jointly released.
\end{enumerate}

\subsection{General setup}
DP was first introduced by \cite{dwork2006calibrating} which defined the concept of an $\epsilon$-indistinguishable RM, now commonly known as $\epsilon$-DP or ``pure" DP RM. Since then, DP as a framework has spawned a massive number of new privacy definitions \citep{desfontaines2019sok}. Because of this, it is often unclear and debated what makes any particular property emblematic of ``DP." Here, we restrict ourselves to the most common privacy definitions and properties associated with the majority of DP implementations \citep{dwork2014algorithmic}. 

DP methods aim to limit the probabilistic influence of any individual's database contribution on sanitized outputs $S(D)$ (again, $S(D)$ could be any summary statistic, parameter estimation, or synthetic microdata sample). In doing so, DP methods heuristically limit what can be inferred about an individual's contribution to an output, regardless of whether they contribute to the database or not, in excess of what an adversary might know a priori. Ideally, whether individuals choose to contribute to a database should not substantively change the overall statistical results. This demonstrates a close connection between DP and robust statistics \citep{Dwork2009,avella2021privacy,slavkovic2021perturbed}. 

Formally, let $D, D'$ be two databases. Let $d_H$ be the Hamming distance between the databases, i.e., the number of elements in the databases that differ:
\begin{equation}
    \label{eq:hamming}
    d_H(D, D') = \#\{ i \in [n] \mid D_i \neq D_i' \}.
\end{equation}
For $d_H$, we say $D,D'$ are adjacent if $d_H(D,D') = 1$. We refer to this case as bounded DP, but we may alternatively consider statistics on databases whose size differ by 1 (known as unbounded DP).

The overall goal of DP is to ensure $S(D)$ and $S(D')$ are close together with high probability when $D$ and $D'$ are adjacent. This is done by parameterizing the distance between $S(D)$ and $S(D')$ with functions of scalar parameters known as privacy loss budgets (PLB), which leads to different DP definitions.  Typically, PLBs are positive, real-valued numbers that capture the trade-off between privacy and data utility; as PLBs increase, more informative statistical results may be released with weaker privacy guarantees. Different definitions have different PLB accounting systems (ex: $\epsilon$ for $\epsilon$-DP, $\rho$ for $\rho$-zCDP): 

\begin{marginnote}
\entry{Privacy loss budget (PLB)}{a scalar parameter that quantifies DP guarantees, with smaller values conferring stronger privacy (ex: $\epsilon$ in $\epsilon$-DP).}
\end{marginnote}

\begin{itemize}
    \item $\epsilon$-DP \citep{dwork2006calibrating}: define the log max-divergence as,
    \begin{equation}
        \label{eq:max_div}
        D_\infty(S(D) \mid \mid S(D')) \triangleq \sup_{B \in \mathcal{F}} \log\left(\frac{\p(S(D) \in B)}{\p(S(D') \in B)}\right).
    \end{equation}
    If $S$ satisfies $\epsilon$-DP for some PLB $\epsilon \in [0, \infty)$, then for all $B \in \mathcal{F}$ and databases, $D, D'$ with $d_H(D, D') = 1$:
    \begin{equation}
        \label{eq:edp}
        \p(S(D) \in B) \leq \p(S(D') \in B) e^\epsilon.
    \end{equation}
    This is equivalent to bounding $D_\infty(S(D) \mid \mid S(D')) \leq \epsilon$ for all adjacent $D, D'$.
    \item $(\epsilon,\delta)$-DP \citep{dwork2006our}: similarly, we can relax $\epsilon$-DP by incorporating a relaxation parameter, $\delta \in [0, 1)$.
    \begin{equation}
        \label{eq:eddp}
        \p(S(D) \in B) \leq \p(S(D') \in B) e^\epsilon + \delta.
    \end{equation}
    \item $\rho$-zCDP \citep{bun2016concentrated}: Define the R\'enyi divergence:
    \begin{equation}
        \label{eq:rzdp}
        D_\alpha(S(D) \mid \mid S(D')) \triangleq \frac{1}{\alpha - 1} \int \log\left( \frac{p(S(D))^\alpha}{p(S(D'))^{\alpha-1}} \right) \ d\mathcal{S},
    \end{equation}
    where $p(\cdot)$ is the density of the mechanism, and the integral is taken over the statistical output space $\mathcal{S}$. Then a RM satisfies $\rho$-zero-concentrated DP (or $\rho$-zCDP) if for all $\alpha \in (1, \infty)$,
    \begin{equation}
        \label{eq:zcdp}
        D_\alpha(S(D) \mid \mid S(D')) \leq \alpha \rho.
    \end{equation}
\end{itemize}

Next, we present some of the nice statistical interpretations of $\epsilon$-DP, the strongest of the three definitions above (in that satisfying $\epsilon$-DP implies satisfying the other two definitions). In the same setup, consider the hypotheses:
$$
H_0: X_1 = x_0, \quad X_1 \neq x_1, \quad x_0, x_1 \in \mathcal{X}.
$$
Note that we assume database rows are exchangeable, in that \textit{any} user's contribution may serve as $X_1$. \cite{wasserman2010statistical} show that if $S(D)$ is an $\epsilon$-DP result, then any procedure for testing $H_0$ above based on $S(D)$ with type I error $\alpha$ has power bounded above by $\alpha e^\epsilon$. Extending this interpretation, \cite{dong2019gaussian} consider a functional analogue of the PLB and its connections to the trade-off between Types I and II errors. Similar testing interpretations exist for $\rho$-zCDP as sub-Gaussian concentration equalities on the log-likelihood ratio for this test. From a Bayesian perspective, if $\pi$ is any prior distribution on the hypotheses, then if $S(D)$ satisfies $\epsilon$-DP:
\begin{equation}
    \label{eq:edp_bayes}
    \frac{\p(H_1 \mid S(D))}{\p(H_0 \mid S(D))} \Big/ \frac{\pi(H_1)}{\pi(H_0)} \in [e^{-\epsilon}, e^{\epsilon}].
\end{equation}
All these interpretations capture the important property that DP only protects against relative disclosure risks. For example, Equation \ref{eq:edp_bayes} suggests that under $\epsilon$-DP, an adversary's prior odds of learning information about someone is similar to their posterior odds. Therefore, the adversary's Bayes factor is close to 1 when the PLB is sufficiently small, suggesting that the $\epsilon$-DP result does little to change the adversary's knowledge.

\subsection{DP Release Mechanisms}

Release mechanisms which satisfy DP rely on randomization to ensure the distance between the two distributions of the output are close. Below is but a small sample of the many possible RMs used to satisfy DP. In this section, we consider properties of a statistic $T(\mathcal{D}) \in \mathcal{T}$ that we aim to release. Central to many different DP definitions is the concept of sensitivity, defined as $\Delta$ where for some norm on $\mathcal{S}$, $||\cdot ||$: 
$$
\Delta \triangleq \sup_{D,D' \in \mathcal{D}, d_H(D,D') = 1} ||T(D) - T(D')||.
$$

\begin{marginnote}
\entry{Sensitivity}{the largest possible change, $\Delta$, in a statistic, $T(D)$, by altering one entry in the database $\mathcal{D}$, as measured by a norm $||\cdot||$.}
\end{marginnote}

Sensitivity captures the worst-case influence of one individual on $T$, which depends on $||\cdot ||$. Notably, the optimal choice of norm for any particular $T(D)$ can be inferred by the geometric properties of the sensitivity space $\mathcal{S}_T$ \citep{awan2020structure}:
$$
\mathcal{S}_T \triangleq \{ T(D) - T(D)' \mid D,D' \in \mathcal{D}, d_H(D, D') = 1 \}.
$$
For many statistics of interest, the sensitivity space (and thus $\Delta$) is bounded by construction. As an example, count data within a single cell (i.e., the number of database users with a particular attribute), has a sensitivity of 1. However, for more complex statistics, $\Delta$ may be unbounded (ex: a parameter in a linear regression). The most common approach to address this problem is to introduce enforced bounds, either by bounding the output space $\mathcal{S}$, database input space $\mathcal{D}$, or the parameter space in a potential model for the data, $\Theta$. This implementation choice has important consequences for validity and consistency of downstream statistical inference (see Section \ref{sec:misspec}). We also note that some (e.g.,  \cite{wang2015privacy,minami2016differential}) use regularity conditions induced from posterior sampling to sidestep the problem of unbounded sensitivity. 

\subsubsection{Primitive Elements}

Most DP algorithms rely on a few primitive components that are subseqently post-processed and composed in complete algorithms. Here, we review the most common primitives. The simplest way to satisfy DP is to add independent noise to $T(D)$, i.e., $S(D) \triangleq T(D) + \gamma$, where $\gamma$ is a random variable with mean 0 and variance that increases as $\Delta$ increases and the PLB decreases. Notable examples for $\epsilon$-DP include the Laplace mechanism \citep{dwork2006our}, its discrete analogue \citep{Ghosh2012}, and the family of $K$-norm mechanisms \citep{hardt2010geometry,awan2020structure}. Examples for $(\epsilon, \delta)$ and $\rho$-zCDP include the Gaussian mechanism \citep{dwork2006our} and its discrete analogue \citep{canonne2020discrete}.
    
Alternatively, we can consider solving an optimization problem based on confidential data while simultaneously satisfying $\epsilon$-DP. This is canonically associated with the $\epsilon$-DP exponential mechanism \citep{mcsherry2007mechanism}, in which a loss function $L: \mathcal{S} \times \mathcal{D} \times [0, \infty]$ is minimized while respecting $\epsilon$-DP. When
\begin{equation}
    \label{eq:expmech_sens}
    \sup_{s \in \mathcal{S}} \sup_{D,D' \in \mathcal{D}} || L(s, D) - L(s, D') || \leq \Delta_L,
\end{equation}
we can satisfy $\epsilon$-DP by releasing one sample from the density
\begin{equation}
    \label{eq:expmech_dens}
    f(s) \propto \exp\left( -\frac{\epsilon}{2\Delta} L(s, D) \right) \nu(s),
\end{equation}
where $\nu(\cdot)$ is a base measure which does not depend on $D$. Notable choices which allow for nice asymptotic properties include the inverse sensitivity mechanism \citep{asi2020} and $K$-norm gradient mechanism \citep{reimherr2019kng}, which are equivalent for some common classes of learning problems.
            
Some optimization problems in statistics and machine learning can be solved by perturbing the input to the problem, i.e.:
\begin{equation}
    \label{eq:obj_pert}
    S(D) = \mathrm{argmin}_{s \in \mathcal{S}} \left[ L_D(s) + \gamma \right],
\end{equation}
where the form of $\gamma$ is chosen based on the problem constraints and the PLB. These have been used in empirical risk minimization \citep{chaudhuri2011differentially}, convex optimization \citep{kifer2012private}, and robust $M$-estimation \citep{slavkovic2021perturbed}, to name a few.

\subsubsection{Post-processing and Composition Techniques}

DP's properties enable flexibility in constructing complex algorithms from primitive building blocks. First, post-processing allows us to construct DP statistics by transforming DP microdata under the same PLB. Second, we can generate DP parameter estimates for two different models and understand their privacy guarantees using sequential composition (e.g., from the same data release $S_1(D)$ with $\epsilon_1$ and $S_2(D)$, $\epsilon_2$, and the total PLB will be cumulative). Third, we can apply a DP method to different sub-populations of interest in a database, and maintain the same privacy guarantees through parallel composition \citep{mcsherry2009privacy}. 

Thus, given the primitives and the properties, there are countless ways to engineer more complex DP algorithms. Here, we highlight common clusters of techniques. First, because the primitive mechanisms depend so heavily on sensitivity, artificial regularity is often induced on $D$ to reduce this sensitivity. While this can be done using SDC techniques (truncation, discretization, clipping, etc.), more advanced methods exploit dimension reduction to effectively reduce the sensitivity of correlated statistics, such as the high-dimensional matrix mechanism for large counting query collections \citep{mckenna2018optimizing} or private PCA for linear dimension reduction \citep{chaudhuri2012near,awan2019benefits}. 
		
For large data sets, subsampling provides a natural way to reduce the effective PLB for different mechanisms (often referred to as ``subsample-and-aggregate" in DP) \citep{nissim2007smooth,li2012sampling}. For example, if $\eta$ * 100\% of a population is subsampled from an $\epsilon$-DP result, then the resulting effective $\epsilon^*$ is $O(\eta \epsilon)$. Natural extensions of subsampling include private bagging and boosting \citep{dwork2010boosting,jordon2019differentially}.

Private synthetic data generation can be viewed as resampling from a model with parameters privately estimated from confidential data. While the regularity introduced by Bayesian priors offers some inherent privacy protections \citep{wang2015privacy}, other approaches involve samples privately weighted by synthetic data utility \citep{snoke2018pmse,vietri2020new}, modeling with Bayesian networks \citep{mckenna2019graphical} and GAN modeling \citep{torkzadehmahani2019dp}. Each of these methods offer different empirical  \citep{bowen2019comparative} benefits.

As an aside, private building blocks can be used to reconstruct most machine learning methods while satisfying DP. As one example, private stochastic gradient descent \citep{song2013stochastic} and its countless variants has allowed for the mass proliferation of DP deep learning methods \citep{boulemtafes2020review}. These methods frequently use $\rho$-zCDP, which has gained popularity in the machine learning community since it relies on Gaussian noise, and learning-theoretic properties of sub-Gaussian distributions form the foundations for statistical learning theory \citep{bousquet2003introduction,vershynin2018high}. We point readers to \cite{vadhan2017complexity} for a review on the sample complexity of DP. 


\section{Data Utility under SDP}
\label{sec:sdp_shared}

In both approaches to SDP, we release sanitized statistics $S(D)$ out into the wild. What happens next? In the previous sections, we discussed the privacy properties of $S(D)$ under SDC and DP independently; now, we consider the data utility properties of arbitrary sanitized outputs $S(D)$, regardless of their privacy semantics. 

``Data utility" is itself ambiguous, so we need to unpack the term. We again let $T(D) \in \mathcal{T}$ be our statistic of interest without any privacy preservation applied (i.e., our ``unsanitized" or confidential statistic). Our goal is to perform inference on a parameter $\theta \in \Theta$. In doing so, we can ask many different questions:

\begin{itemize}
    \item \textbf{Data-based utility}: how close is my sanitized output $S(D)$ to the confidential output $T(D)$?
    \item \textbf{Comparative inferential utility}: how close is a sanitized estimator $\hat\theta(S(D))$ to a confidential estimator $\hat\theta(T(D))$?
    \item \textbf{Estimator inferential utility}: how is my uncertainty for $\theta$ using $\hat\theta(S(D))$ different from my uncertainty for $\theta$ using $\hat\theta(T(D))$?
\end{itemize}

Our ability to address these questions depends on whether we are designing the RM $S$ (e.g., release a consistent and asymptotically unbiased sanitized parameter estimate), or adjusting for the effect of RM $S$ which we did not choose (e.g., adjust the length of the confidence interval given the sanitized statistic). When we design a RM for a specific inferential task, all three of these should yield the same relative comparisons between estimators (i.e., if a mechanism offers better data-based utility, it also offers better estimator inferential utility). However, when we adjust for an existing RM, these utility definitions do not offer the same relative comparisons between RMs, and can even be conflicting! As an example from $\epsilon$-DP count data, the Geometric mechanism \citep{Ghosh2012} can optimize data-based utility, but it requires post-processing that is sub-optimal for estimator inferential utility on binomial data \citep{awan2018differentially}. Therefore, we need to address these two problem classes differently.

Here, we express the design and adjustment problems as two different minimax estimation problems (though we could easily pick another loss aggregating convention) \citep{slavkovic2019statistical}. Suppose we want to minimize some loss function $L: \Theta \times \Theta \mapsto \mathbb{R}^+$ in the worst-case scenario over a space of possible data generating distributions $\mathcal{P}$ indexed by $\p \in \mathcal{P}$. For any release mechanisms $S(D)$, this requires us to think about the marginal distributions for $S(D)$ for a given data generating distribution $\p \in \mathcal{P}$, i.e.:
\begin{equation}
    \label{eq:marg_p}
    \mathcal{M}_S(\p) =  \sum_{D \in \mathcal{D}}  \mathrm{Pr}(S(D) \mid D) \ \p_\theta(D) 
\end{equation}

From the design perspective, we are given a space of RMs $\mathcal{Q}$ which satisfies some privacy guarantee. Our goal is to find the optimal RM $S^* \in \mathcal{Q}$ and estimator $\hat{\theta}_{\mathrm{Design}}(S^*(D))$ that satisfies:
\begin{equation}
    \label{eq:design_opt_est}
    \hat{\theta}_{\mathrm{Design}} = \argmin_{\tilde{\theta}, S \in \mathcal{Q}} \max_{\p \in \mathcal{P}} \E_{\mathcal{M}_S(\p)}[L(\tilde{\theta}(S(D)), \theta)].
\end{equation}
This problem has been analyzed in the local DP setting \citep{duchi2018minimax} and similarly in central DP \citep{smith2011privacy}. Alternatively, suppose we are only given a sample $S(D)$ from an RM we did not design ourselves. Then our inference problem requires us to find the optimal adjusted estimator $\hat{\theta}_{\mathrm{Adjust}}(S(D))$ that satisfies:
\begin{equation}
    \label{eq:adj_opt_est}
    \hat{\theta}_{\mathrm{Adjust}} = \argmin_{\tilde{\theta} } \max_{\p \in \mathcal{P}} \E_{\mathcal{M}_S(\p)}[L(\tilde{\theta}(S(D)), \theta)].
\end{equation}

Regardless of whether we choose $S$ or not, statistical inference requires that we account for the transformation $S(D)$, meaning we CANNOT treat inference given $T(D)$ the same as inference given $S(D)$, as the two variables have entirely different sampling distributions; on a related issues of approximating sanitized sampling distributions, see \cite{wang2018approxdist}. This is true for all SDP methods, those from SDC and DP. Not only can the distribution of $S(D) \mid D$ introduced randomized errors due to privacy, the sample spaces of $S(D)$ and $T(D)$ can be entirely different, even for SDC methods involving no randomization. This demonstrates that the de facto practice of naively substituting $T(D)$ with $S(D)$ can produce invalid statistical inferences, with incorrect interpretations of significance, coverage, or other properties of statistical estimators; e.g., for these in a network setting see \citep{karwa2016,Karwa2017}.

\subsection{Specific utility and the design approach} 

First, we consider the design problem, in which our goal is to perform inference for $\theta \in \Theta$ and design a valid estimator $\hat{\theta} = S(D)$, where the RM satisfies some privacy guarantees. In the SDC literature, data utility is frequently quantified by measures that capture statistical information lost due to $S$ \citep{hundepool2012statistical}. In the DP literature, the evaluation of randomized mechanisms relies on concentration inequality results to bound probabilistic distances between $S(D)$ and $T(D)$, or equivalently $\hat\theta(S(D))$ and $\hat\theta(T(D))$ \citep{boucheron2013concentration}. Under consistency or other oracle assumptions, these will give us estimator inferential utility measures as well. 

Focusing on uncertainty quantification directly offers a few advantages. First, we can design optimal estimators based on the degree to which they specifically influence our statistical uncertainty. Examples include power and sample size analysis for experimental data \citep{Vu2009}, confidence interval width \citep{karwa2017finite}, the power of finite-sample hypothesis testing procedures \citep{awan2018differentially}, and asymptotically correct inference from central limit theorem approximations \citep{awan2019benefits}.
Second, these procedures are more user-friendly, as they account for uncertainty in $\hat\theta$ due to privacy preservation. When we strictly measure how close $\hat{\theta}(S(D))$ is to $\hat{\theta}(T(D))$, we cannot draw the same conclusions, because such a comparison does not account for other sources of error in the data generating process.

\subsection{General utility and the adjustment approach}

Alternatively, we consider the adjustment problem, in which we must account for a RM we did not design specifically for our inferential problem. This is the setting most often associated with private synthetic microdata or collections of sanitized statistics, suggesting different kinds of utility measures for general purpose inference and inference on specific tasks \citep{snoke2018general,arnold2020really}. 

Importantly, different methods for generating $S(D)$ may be compatible or incompatible with different probability models for $\theta$. For example, if we generate sanitized estimates of sufficient statistics for $\theta$, then we would say this model is compatible with the RM because we can account for measurement error in a way that still produces asymptotically consistent statistics (see \cite{foulds2016theory} for an example in Bayesian inference). However, if this is not the case, i.e., if the confidential target of our private statistics $T(D)$ is not sufficient for the model, there are certain inferences we cannot perform at all.
 
For inference on general purpose data, we need to characterize the likelihood of $S(D)$ given $\theta$ by integrating out the confidential data. This can be done from the frequentist perspective, i.e.,:
\begin{equation}
    \label{eq:freq_likelihood_sd}
    \p_\theta(S(D)) = \sum_{\mathcal{D}} \p(S(D) \mid D) \ \p_{\theta}(D),
\end{equation}
or from the Bayesian perspective, i.e., with prior $\pi(\theta)$,
\begin{equation}
    \label{eq:bayes_post_sd}
    \p(\theta \mid S(D)) \propto \pi(\theta) \sum_{\mathcal{D}} \p(S(D) \mid D) \p(D \mid \theta).
\end{equation}
Because of this necessity, DP offers a particular advantage over SDC. If $S$ satisfies DP, then the privacy mechanism is transparent (i.e., the form of measurement errors are publicly known), and the problem reduces to a classical ``error-in-variables" problem. For common models, we can readily rely on techniques from the existing measurement error literature, such as techniques based on generalized linear models and estimating equations \citep{tsiatis2006semiparametric,Carroll2006,hardin2002generalized}. 

 Still, incorporating these errors is easier said than done, as the integration in Equations \ref{eq:freq_likelihood_sd} and \ref{eq:bayes_post_sd} can be quite computationally difficult. In some cases, connections between approximate Bayesian computation \citep{beaumont2019approximate} and inference on noisy estimates can be used for posterior inference. \cite{Fearnhead2012} showed that exact inference from perturbed statistics uses the same inferential sampling procedure as ABC with normal summary statistics. This allows \cite{gong2019exact} and \cite{Seeman2020} to produce valid inference from statistical results. Note that different privacy mechanisms are more or less amenable to probability models, which we will see in the next section.

\subsection{Statistical properties of RMs}
\label{sec:sdp_count_mechs}

In this section, we compare a few different DP mechanisms for counting queries and discuss their statistical properties; specifically, we discuss how these choices for $S(D)$ affect the ease of downstream inference through probability models. Suppose we are interested in releasing a count of events $T(D) \in \mathbb{Z}^+$, in which our sensitivity $\Delta$ is 1. We consider different ways of releasing $T(D)$ satisfying different DP formalisms:

\begin{enumerate}[noitemsep]
    \item \texttt{DiscLaplace}: Discrete Laplace mechanism for $\epsilon$-DP \citep{Ghosh2012}:
    \begin{equation}
        \label{eq:disc_lap}
        S(D) = T(D) + \varepsilon, \varepsilon \sim \mathrm{DiscreteLaplace}(\epsilon^{-1}).
    \end{equation}
    \item \texttt{DiscGaussian}: Discrete Laplace mechanism for $\rho$-zCDP \citep{canonne2020discrete}:
    \begin{equation}
        \label{eq:disc_gauss}
    S(D) = T(D) + \varepsilon, \varepsilon \sim \mathrm{DiscreteGaussian}(\rho^{-2}).
    \end{equation}
    \item \texttt{Exponential}: Exponential mechanism for $\epsilon$-DP \citep{mcsherry2007mechanism}:
    \begin{equation}
        \label{eq:expmech_disc}
    \p(S(D) = k) \propto \exp\left( -\frac{\epsilon}{2} |k - T(D)| \right) \mathbbm{1}\left\{ k \in \{ 0, 1, \dots, n \} \right\}.
    \end{equation}
    \item \texttt{RandResp}: Randomized Response for local $\epsilon$-DP:
    \begin{equation}
        \label{eq:rand_resp}
        S(D) = \sum_{i=1}^n \mathrm{RR}(X_i), \quad \begin{cases}
        \p(\mathrm{RR}(X_i) = X_i) = \frac{\exp(\epsilon)}{1 + \exp(\epsilon)} \\
        \p(\mathrm{RR}(X_i) = 1 - X_i) = \frac{1}{1 + \exp(\epsilon)}.
        \end{cases}
    \end{equation}
\end{enumerate}
Note that not all privacy guarantees are the same: local $\epsilon$-DP is stronger than $\epsilon$-DP, which is stronger than $\rho$-zCDP. Furthermore, each mechanism has different statistical properties, which we summarize in Table \ref{tab:disc_mechanisms}
and describe here:
\begin{enumerate}
    \item \textbf{Error independence}: can randomized errors due to privacy be expressed as a perturbation, where for some norm $|| \cdot ||$, $||S(D) - T(D)|| \indep D$?
    \item \textbf{Unbiased}: does the mechanism introduce bias into the estimate of the confidential data, i.e., does $\E[S(D)] = T(D) \ ?$
    \item \textbf{Mode unbiased}: is the maximum likelihood output of the mechanism the confidential response? i.e., is it true that:
    \begin{equation}
        \label{eq:mech_mode_unbiased}
        \max_{S^* \in \mathcal{S}} \p(S(D) = S^*) = T(D) \ ? 
    \end{equation}
    (NB: result assumes the existence of a PMF for $S(D)$, with analogous results for arbitrary measures or densities). 
    \item \textbf{Domain-constrained}: does the space of $T(D)$ conform to the space of $S(D)$? i.e., is it true that for all $B \in \mathcal{F}$:
    \begin{equation}
        \label{eq:mech_domain_constrained}
        \p(T(D) \in B) = 0 \implies \p(S(D) \in B) = 0 \ ?
    \end{equation}
\end{enumerate}

\begin{table}
    \centering
    \begin{tabular}{|c|c|c|c|c|}
        \hline
        \textbf{Mechanism} & DiscLaplace & DiscGaussian & Exponential & RandResp \\
        \hline 
        \textbf{Trust model} & Central & Central & Central & Local \\
        \hline 
        \textbf{Formalism} & $\epsilon$-DP &  $\rho$-zCDP &  $\epsilon$-DP  &  $\epsilon$-LDP \\
        \hline
        \textbf{ErrorDist} $\indep D$? & Yes & Yes & No & No \\
        \hline
        \textbf{Unbiased?} & Yes & Yes & No & No \\
        \hline 
        \textbf{Mode unbiased?} & Yes & Yes & Yes & No \\
        \hline
        \textbf{Domain-constrained?} & No & No & Yes & Yes \\
        \hline 
    \end{tabular}
    \caption{Counting mechanisms and their statistical properties}
    \label{tab:disc_mechanisms}
\end{table}

Note that we could post-process either \texttt{DiscLaplace} or \texttt{DiscGaussian} to restrict the domain, i.e.:
\begin{equation}
    \label{eq:mech_count_post}
    S_{\mathrm{post}}(D) = \begin{cases}
    0 & T(D) + \varepsilon < 0 \\
    n & T(D) + \varepsilon > n \\
    T(D) + \varepsilon & \mathrm{Otherwise.}
    \end{cases}
\end{equation}
This post-processing transformation, proposed in \cite{Ghosh2012}, offers a uniform improvement in utility as measured by the distance between $S(D)$ and $T(D)$, i.e.:
\begin{equation}
    \label{eq:post_loss_util}
    \E[|S_{\mathrm{post}}(D) - T(D)|] \leq \E[|S(D) - T(D)|].
\end{equation}
However, $S_{\mathrm{post}}(D)$ is no longer unbiased, and the errors are now data-dependent. This demonstrates that post-processing changes the statistical properties of RMs, and improving utility compared to confidential results can have unintended consequences for the statistical properties of these estimators. In fact, post-processing can degrade the power of resulting statistical inferences, sometimes uniformly \citep{Seeman2020, seeman2022formal}. Therefore, it is essential that we consider which mechanisms are amenable to downstream inference and which make it prohibitively difficult or computationally expensive. 

\subsection{Risk and model misspecification} 
\label{sec:misspec}
As one last important caveat, we remind ourselves that whenever we make modeling assumptions, there is always the potential to be wrong. SDP introduces new opportunities for different kinds of misspecification that we briefly discuss here. 

SDP relies on the properties of the database schema $\mathcal{D}$ being correctly specified. When this is not the case, SDP risk and utility can both suffer. Unanticipated records falling outside the expected schema could result in weaker DRMs (e.g., negative counts being dropped); if those records are systematically excluded due to processing errors, then we could be subject to an unknown form of unaccounted missing data. As another example, we may incorrectly specify the sensitivity of a statistic $T(D)$, meaning our privacy guarantees are realized at a larger PLB than intended. 

SDP risk measures may be based on implicit assumptions on independence between records which may not hold in practice. For an example with DP, the framework is colloquially seen as demonstrating ``robustness to arbitrary side information," as formalized in \cite{kasiviswanathan2008note}. However, when DP is treated as a property of probability distributions, as in Pufferfish privacy by  \cite{kifer2014pufferfish}, we only maintain $\epsilon$-DP style privacy semantics when the entries of the database are independent. In other words, when there is dependence between database records, $\epsilon$-DP guarantees may be degraded \citep{liu2016dependence}. This has motivated methods for the private analysis of correlated data \citep{song2017pufferfish,Karwa2017,seeman2022formal}. 

\section{The Fable of the Westerosi Census}
\label{sec:case_study}

Many of the core research problems in SDP rely on translating statistical notions into practical commitments to privacy protections and data utility goals. Here, we do not focus on the newest, nor the most advanced mechanisms by modern publishing standards. Instead, the goal of this section is to showcase that there is complex interplay between different privacy formalisms, the underlying data structure, and data generating assumptions, all of which affect risk and utility, that is valid statistical inference. To the end, we turn to a fabulistic case study on our data.



We consider a dataset based on the population of fictional characters from the world of Westeros in the fantasy series ``Game of Thrones" (GoT) \citep{martin2011game}; data were gathered by mining the text of the fan-written Wikipedia, ``A Wiki of Ice and Fire" \citep{awoiaf,tum2019song}. Variables are described in Table \ref{tab:got_data_desc}. Note that in working with this dataset, we do not intend to make light of the very real harms caused by privacy violations. Instead, we use a dataset where such harms are impossible by construction, and the worst possible harms for readers are minor story spoilers. And to the best of our knowledge, it is impossible to violate the privacy of a fictional character. 

Our data curator, Lord Varys (henceforth LV) is tasked with conducting a census of the citizens of Westeros, in order to count and report which citizes have survived the politically tumultuous events of GoT. However, he is concerned about an adversary, Littlefinger (henceforth LF), learning information about vulnerable members of the royal family whose data may be contained in the Census. For this case study, we consider whether or not a character survives the events of GoT to be a sensitive attribute. We demonstrate, how LV's assessment of risk and utility changes in different scenarios. 

\begin{table}[!htbp]
    \centering
    \begin{tabular}{|c|c|c|}
        \hline 
        \textbf{Variable} & \textbf{Kind} & \textbf{Description} \\
        \hline
        \texttt{title} & Nominal & Title (ex: Ser, Lord) (261 levels) \\
        \hline 
        \texttt{title\_reduced} & Nominal & Major Title Categories (11 levels) \\
        \hline 
        \texttt{culture} & Nominal & Culture (ex: Dornish, Braavosi, Dothraki) (35 levels) \\
        \hline 
        \texttt{culture\_reduced} & Nominal & Major Culture Categories (14 levels) \\
        \hline 
        \texttt{house} & Nominal & House (ex: Stark, Lannister, Tyrell) (326 levels) \\
        \hline 
        \texttt{house\_reduced} & Nominal & Major Houses (27 levels) \\
        \hline 
        \texttt{gender} & Binary & Male or Female \\
        \hline 
        \texttt{nobility} & Binary & Noble or Peasant \\
        \hline
        \texttt{alive} & Binary & Alive or Dead \\
        \hline 
    \end{tabular}
    \caption{Data variable description}
    \label{tab:got_data_desc}
\end{table}

\subsection{Counting queries with SDC}

First, LV considers different contingency tables aggregated based on different quasi-identifiers, (i.e., combinations of different nominal and binary variables). For example, he could create a \texttt{Culture+Nobility} table, which lists the number of nobles and peasants from every culture. He soon realizes that too many respondents have unique combinations of titles, cultures, and/or houses; for example, there is only one Dornish princess in the data, and LF could learn attributes about the Dornish princess if the database is released as-is. Therefore, to address these issues, he creates simplified versions of all these categorical random variables, only keeping categories with at least 10 respondents (as shown in Table \ref{tab:got_data_desc}) and grouping all others into a separate category. We refer to these as the reduced versions of these nominal variables, e.g., \texttt{CultureReduced}.

\begin{figure}[!htbp]
    \centering
    \includegraphics[width=\textwidth]{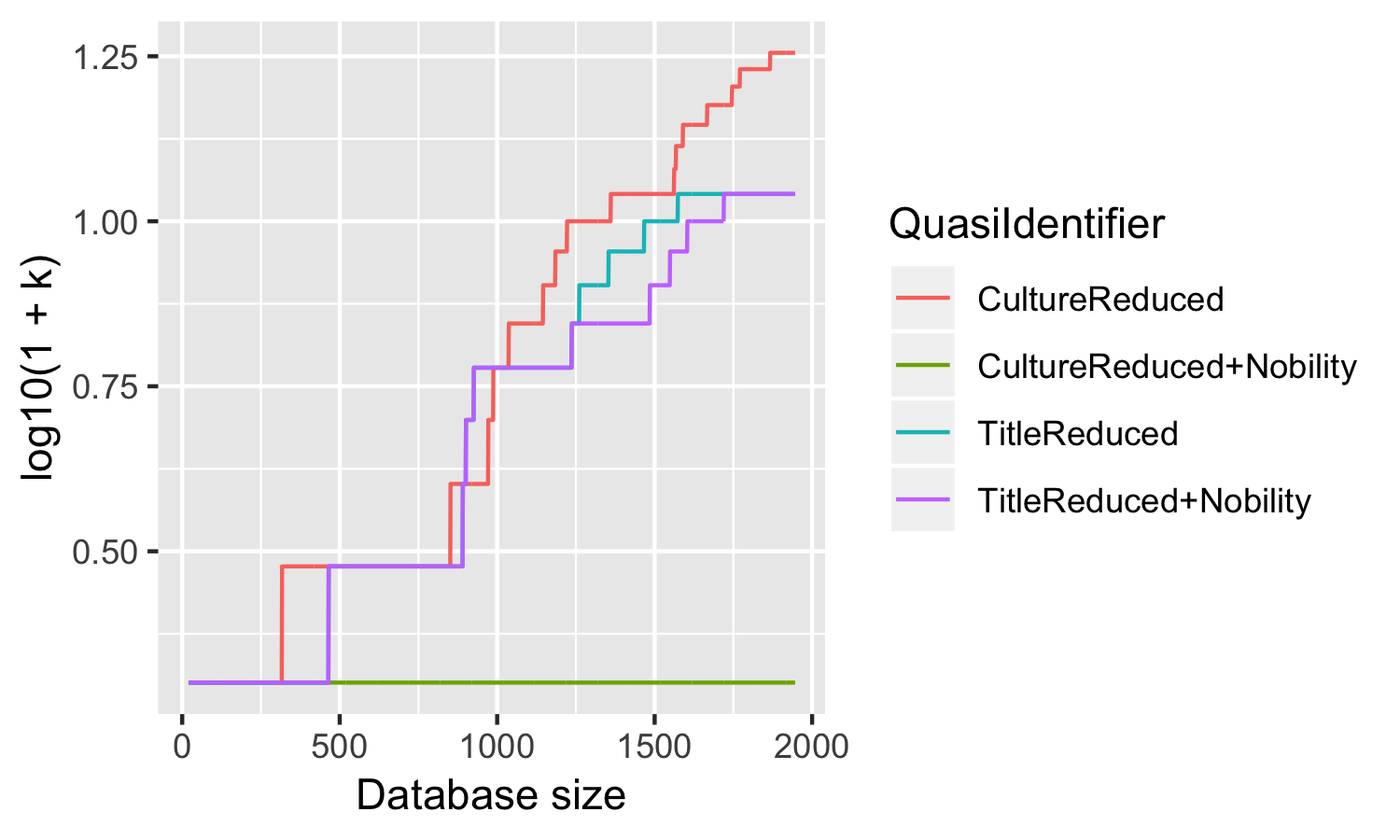}
    \includegraphics[width=\textwidth]{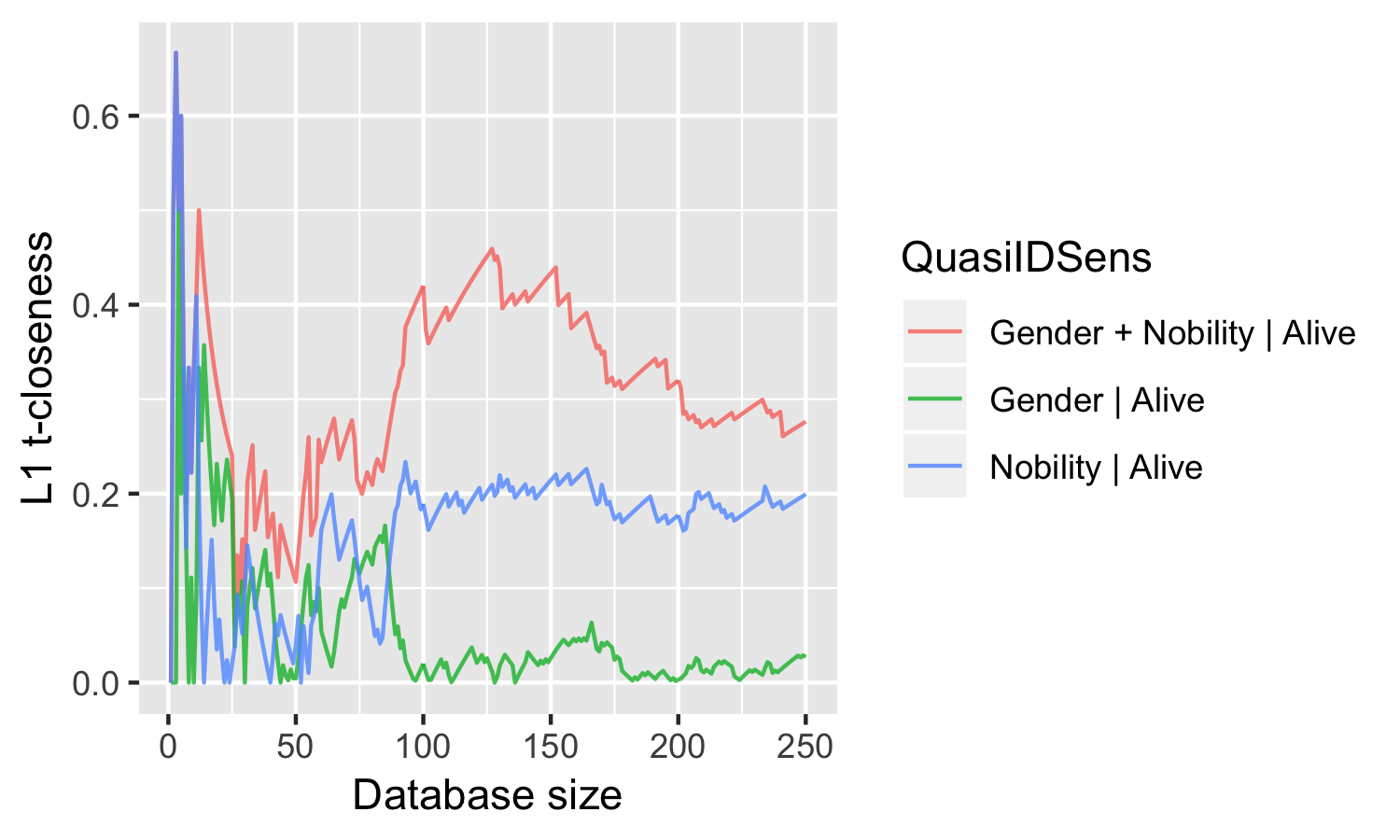}
    \caption{GoT data, (top) $k$-anonymity and (bottom) $t$-closeness  risk measures for different aggregated counts by query (quasi-identifiers) and database size.}
    \label{fig:count_sdc_risk}
\end{figure}

Next, LV considers two SDC risk measures for different contingency tables. First, he looks at $k$-anonymity (Eq. \ref{eq:k_anon}), which captures the smallest number of census respondents with any given quasi-identifier; LV desires larger $k$ values for stronger privacy protections. In the top panel of Figure \ref{fig:count_sdc_risk}, he plots the $k$-anonymity of the first $s$ rows on the $y$-axis with $s$ on the $x$-axis, and sees that different combinations of quasi-identifiers offer different protections; \texttt{CultureReduced} generally has the best $k$-anonymity guarantees, and these increase as the database size increases. However, LV notices that some of the cultures, even after aggregating, have only one noble, meaning that using \texttt{CultureReduced+Nobility} offers $1$-anonymous privacy, regardless of the database size. This means for any database size, if LV releases \texttt{CultureReduced+Nobility}, LF will be able to reidentify anyone with a unique reduced culture and nobility status, and at least one such person exists in the database.

LV is also concerned about what LF might do if he learns which kinds of people are most affected by the events in GoT. So he measures $t$-closeness based on survival, or the largest difference in survival rate between any quasi-identifying group and the overall sample. In the bottom panel of Figure \ref{fig:count_sdc_risk}, he sees that the $t$-closeness is small for \texttt{Gender} (blue line); that means he can release information about the survival rates for men and women in Westeros, and LF is not likely to learn much about anyone from the Census's survival rate simply because of their gender. However, the $t$-closeness is larger for \texttt{Gender+Nobility}; this raises concerns for LV, because he is concerned LF might learn about the probability that someone in the database, like the Dornish princess, is alive or not. However, $t$-closeness can sometimes capture population-level effects; so maybe, LF would learn about the difference in survival rates between noble females and the whole population, regardless of whether the Dornish princess completed the Census of Westeros or not! 

\begin{figure}[!htbp]
    \centering
    \includegraphics[width=\textwidth]{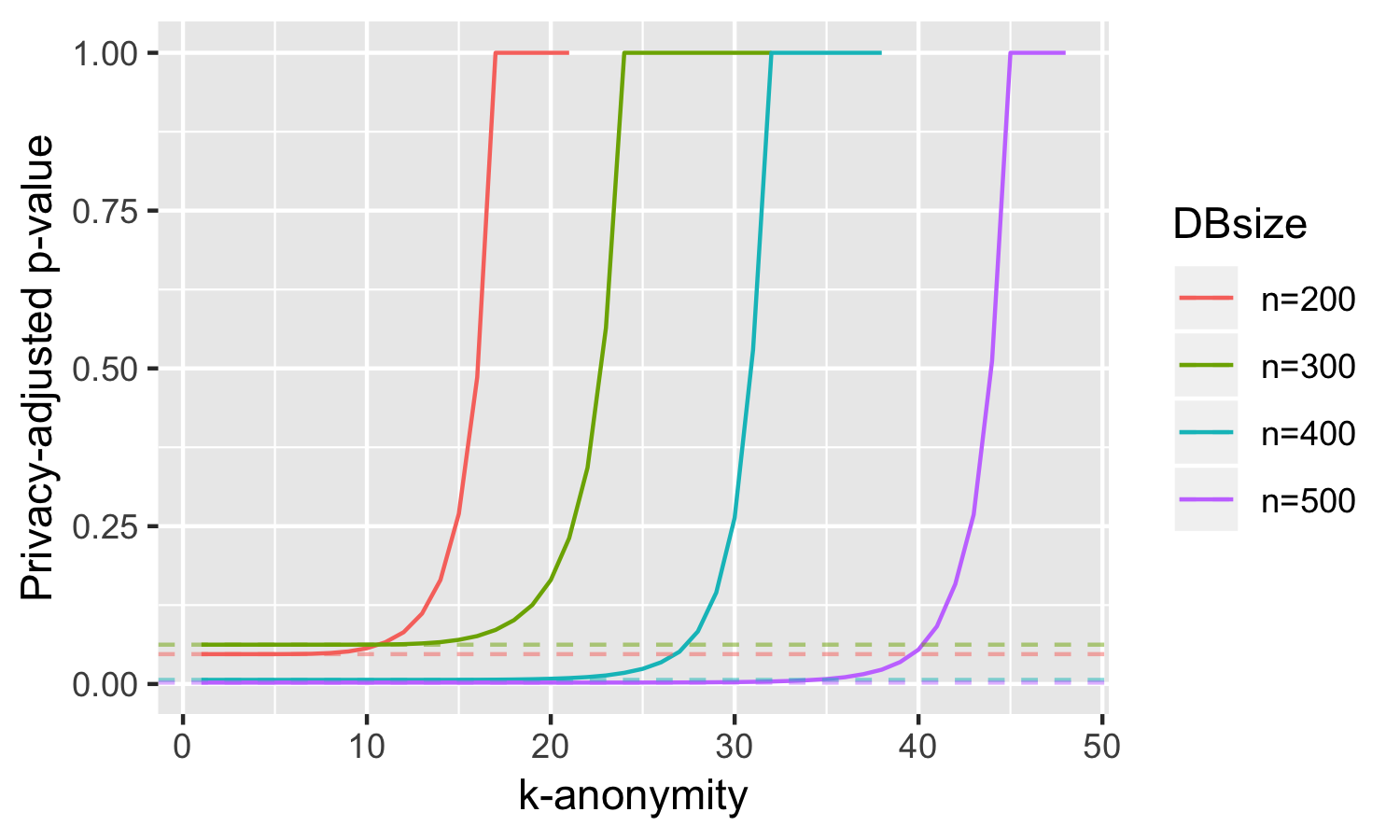}
    \caption{GoT data, privacy-corrected $p$-values for Fisher's exact test ($H_0=$ nobles at least as likely as peasents to survive) from $k$-anonymous tabular data at different database sizes and $k$ values.}
    \label{fig:count_sdc_util_exacttest}
\end{figure}

LV decides he will release a 2x2 contingency table of nobility vs. survival with different $k$-anonymity protections. He wants to perform inference for the null hypothesis $H_0:$ nobles are at least as likely as peasants to survive GoT, with $H_1$: otherwise. Under the null, the number of surviving nobles follows a Hypergeometric distribution. Normally, LV would simulate from this distribution and numerically estimate the test statistic, since he wants to extract as much information as possible. But LV knows his inference is affected by his choice of $k$, and while others may ignore that fact, he conditions on the database being $k$-anonymous for different $k$ values by rejecting simulated statistics that violate $k$-anonymity. LV then uses these reduced samples to calculate the $p$-value, on the $y$-axis in Figure \ref{fig:count_sdc_util_exacttest}. In this risk-utility plot, the horizontal lines refer to the non-private $p$-values, and the solid lines are the estimated $p$-values (on the $y$-axis) at different $k$s (on the $x$-axis). As $k$ increases, LV loses power to detect differences in the survival rate of nobles and peasants; for example, if $n=300$ and $k=20$, we fail to detect such a difference at a Type I error of $.10$. This demonstrates that SDC measures affect statistically valid inference, even when no randomization occurs. 

\subsection{Counting queries with DP}

Because of LV's concerns, including that he cannot release much data under the SDC framework since the risk with \texttt{Nobility} is high, and the fact that he really cannot be sure what LF may already know, he decides to use bounded $\epsilon$-DP methods for counting queries. He first plans to release the counts of the surviving and dead, aggregated by \texttt{CultureRedued+Gender+Nobility}, and with discrete Laplace noise added to each count (Equation \ref{eq:disc_lap}). This way, LF will not be able to learn more about individuals in the database, because the possible releases where any one respondent survived GoT or not are close together with high probability. But this has LV wondering: how much could LF know already, and how much could he stand to learn from using LV's DP census results? LV decides to analyze two different scenarios: first, maybe LF randomly guesses (i.e. flips a fair coin) to determine whether someone has survived GoT or not. Second, maybe LF knows the true confidential proportion of people who have survived GoT, in which case he is a more informed adversary (for reference, around 25\% of characters in GoT die). Using these two priors, LV calculates the posterior risk that LF learns whether the last person in any quasi-identifying cell (\texttt{CultureRedued+Gender+Nobility}) survived or died, in the worst-case scenario where LF knows all but the last entry in any table cell.

\begin{figure}[!htbp]
    \centering
    \includegraphics[width=\textwidth]{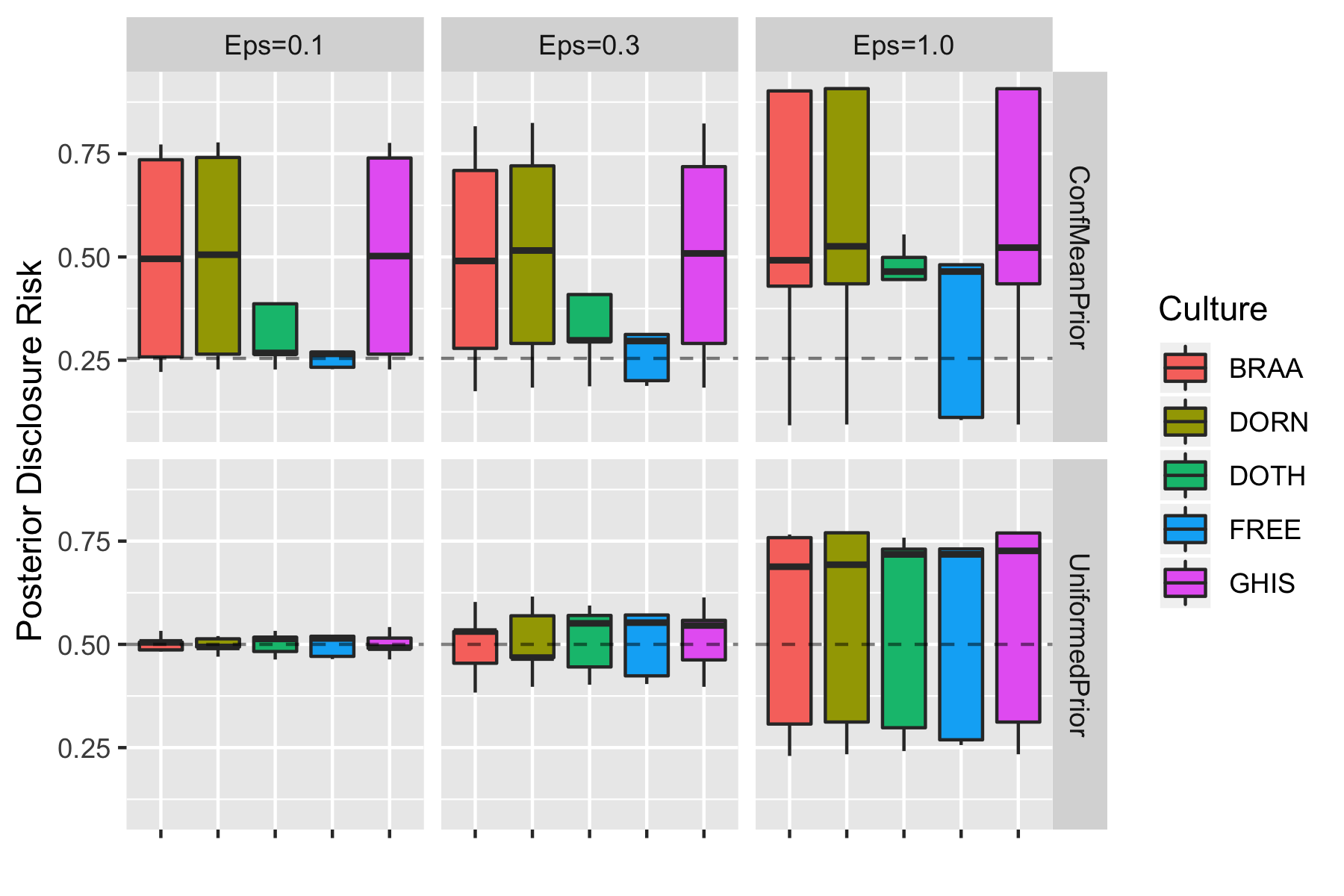}
    \caption{Posterior disclosure risks for individual GoT characters by prior assumptions, culture, and PLB.}
    \label{fig:count_fp_risk_post}
\end{figure}

In Figure \ref{fig:count_fp_risk_post}, LV plots the posterior disclosure risk on the $y$-axis and looks that these posterior disclosure risks. First, the worst-case disclosure risks are not the same for all cultures. Certain minority cultures under the informed prior (top row of panels), such as Braavosi (BRAA) and Dornishmen (DORN), are more likely to be reidentified at small PLBs (e.g., $\epsilon=0.1$) than people of larger cultures, such as Free folk (FREE). Second, we see that prior assumptions can change how these posterior disclosure risks are distributed amongst the populations; if LF has a good prior, he could potentially learn more about whether the Dornish princess survives GoT than, say, someome from the Free folk. This visually demonstrates that even though the worst-case privacy guarantee in $\epsilon$-DP applies to everyone, not everyone has the same posterior disclosure risks.

\begin{figure}[!htbp]
    \centering
    \includegraphics[width=\textwidth]{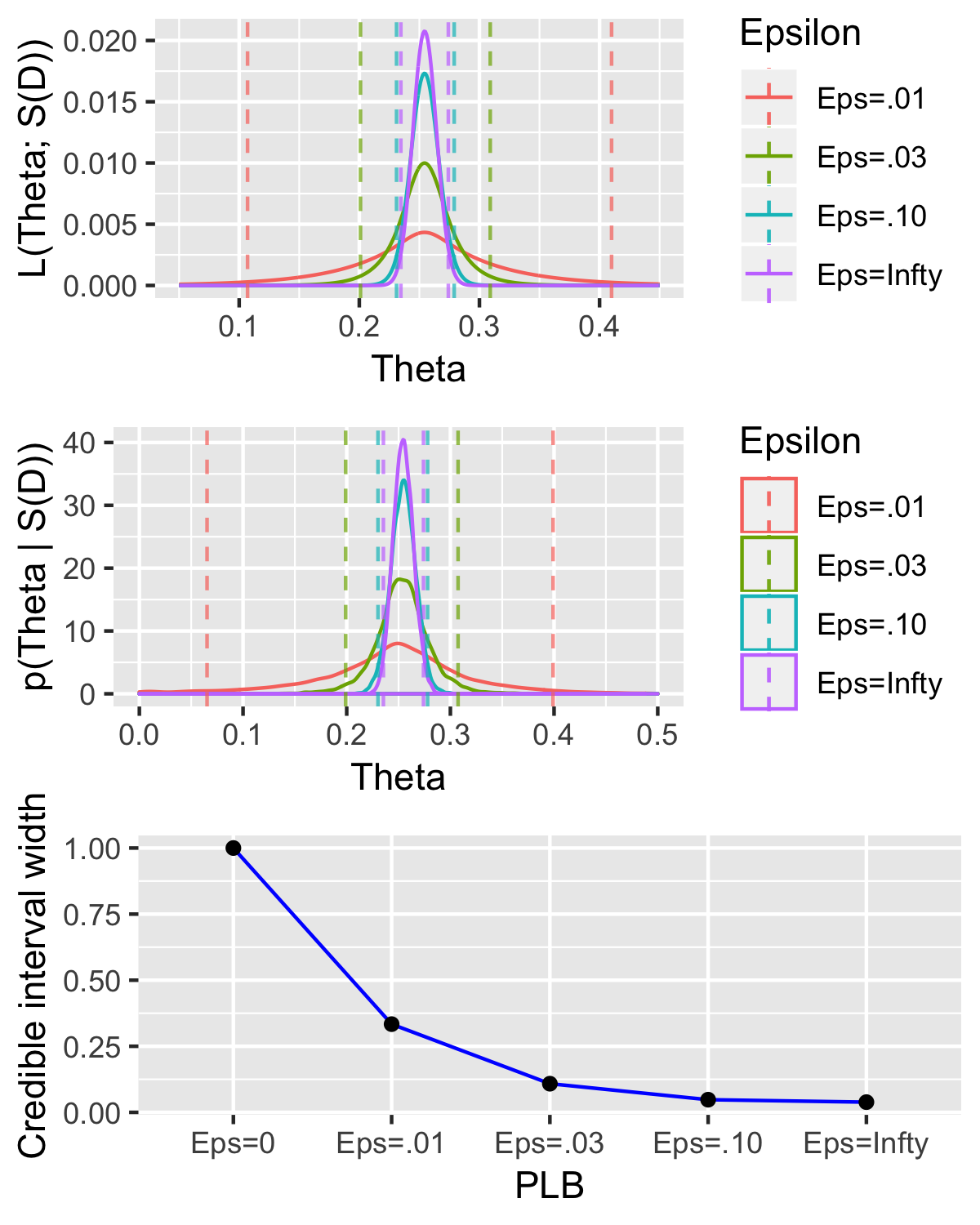}
    \caption{(top) Naive Inference: Likelihood of $\theta$ given $S(D) = T(D)$ under $T(D) \sim \mathrm{Binomial}(n, \theta)$ and associated confidence intervals at different PLBs. (middle) Adjusted Inference:  Posterior distributions and associated credible intervals for $\theta \mid S(D)$ at different PLBs. (bottom) Risk-utility curve for private posterior inference on the GoT survival rate.}
    \label{fig:count_fp_util_lik}
\end{figure}

LV realizes from that analysis that if he wants to release the overall survival rate (or equivalently, the death rate), he needs to sanitize it, even though it is just a summary statistic. But LV still wants to do inference on what this value could be, requiring him to account for additional noise due to privacy. In Figure \ref{fig:count_fp_util_lik}, LV plots three important figures at varying PLBs, from $\epsilon = .01$ (stronger privacy protections) to $\epsilon = \infty$ (non-private release). The upper figure shows the likelihood of the death rate where we assume $S(D) = T(D)$ (chosen for comparison with non-private inference). For this figure, LV also plots 95\% confidence intervals based on the sampling distribution of $S(D)$ (dashed line), which increase in size as the PLB decreases. For some PLBs, like $\epsilon=.10$, the errors due to privacy are dominated by errors due to sampling. For other PLBs, like $\epsilon=.01$, the opposite is true. This essential information can only be inferred by comparing the probability model to the errors due to privacy.

Most importantly, the middle panel of Figure \ref{fig:count_fp_util_lik} tells LV posterior credible intervals (dashed lines) for $\theta \mid S(D)$ under the Jeffery's prior $\theta \sim \mathrm{Beta}(.5, .5)$. Because LV properly accounted for errors due to privacy in his inference, his resulting credible interval increases in width as $\epsilon$ decreases while providing exact statistical coverage. This is NOT true in the case when we naively substitute $T(D)$ with $S(D)$, demonstrating once again the essential nature of statistically valid inference for sanitized results. Moreover, LV plotted this credible interval length on the $y$ axis with the PLB on the $x$-axis. This allows him to visualize the trade-off between privacy and utility and choose a PLB. LV sees that even the non-private result, $\epsilon = \infty$, has a small amount of uncertainty. By sacrificing a little data utility by making his credible interval wider for the number of dead in GoT, LV can help protect the citizens of Westeros from LF, regardless of whether it was the Dornish princess or someone else who completed the Census of Westeros. Huzzah!

\begin{textbox}[h]\section{The Morals of Our Fable}
\begin{itemize}
	\item SDC measures can provide strong or weak privacy protections that scale differently with database sizes, and they may sometimes capture population-level effects that exist regardless of whose data contributed to the population-level inferences.
	\item All SDP methods affect downstream inference, even SDC methods that do not involve randomized noise.
	\item Although DP methods offer worst-case relative privacy guarantees, the posterior disclosure risks look different for different members of the database and different adversarial prior assumptions.
	\item DP necessitates adjusting downstream inferences for errors due to privacy, which requires analyzing the interaction of probability models with statistical measures.
\end{itemize}
\end{textbox}

\newpage 
\section{Discussion \& Future Directions}
\label{sec:disc}

SDC and DP are schools of thought that frame the underlying problems of data privacy in different ways, as there are theoretical and empirical pros and cons to both approaches. Moreover, choosing to frame data privacy problems from one perspective or the other induces trade-offs that cannot always be quantitatively captured; these may be better solved by the legal and normative literatures on data privacy. In this section, we discuss the high-level differences between these approaches along a few key dimensions.

\subsection{Comparing SDC and DP: Quantifying Risk}

SDC and DP rely on measures of privacy risk with different conceptual trade-offs, as discussed in Section \ref{sec:prob_form}. How can we tell which framework is better suited for a particular use case? For that, we need to think about the gaps left from either perspective. For SDC, the main question left unanswered is whether bounds on the DRMs allow for resilience against other kinds of inference attacks; e.g., by using database reconstruction attacks on public Census data, the Census Bureau was able to identify real vulnerabilities in their previous SDC methodology \citep{garfinkel2019understanding}. For DP, the main question left unanswered is how to sociologically interpret PLBs, requiring us to reason about worst-case adversaries, database pairs, and disclosure scenarios. This shift in language can make it hard to express privacy concerns in terms of PLBs \citep{cummings2021need}. Making PLBs more interpretable, though, often requires further assumptions. For example, using the Bayesian formulation of $\epsilon$-DP, we can make prior assumptions and calculate different posterior disclosure risks under such protections \citep{Reiter2012}. Such measures offer more interpretability to practitioners at the expense of no longer providing worst-case guarantees.

Interpreting SDC and DP guarantees depends on many different considerations. How amenable is our output statistic space to privacy-preserving inference? How sociologically sensitive are the attributes about units we observe? How large is our database? What kinds of variability do we expect in the attribute responses? In Sections \ref{sec:sdp_shared} and \ref{sec:case_study}, we only began to scratch the surface of answering these questions with respect to database size and query selection. Still, we demonstrated that answers to these questions change the social calculus of how we aim to quantify and limit privacy risks.

Future work could address an alternative approach, where the unit of analysis is neither a single database (as in SDC) nor an entire schema (as in DP), but instead a restricted set of database pairs within the schema. This approach, sitting somewhere between SDC and DP, could prove useful by making the DP-style "worst-case-scenario" analysis for DP less extreme while still providing more robustness to database reconstruction than SDC. Examples of early work in this area include \citep{kifer2014pufferfish,song2017pufferfish,seeman2022formal}.

\subsection{Comparing SDC and DP: Quantifying Utility and Uncertainty}

The role of data dependence is another distinguishing factor in comparing SDC and DP. Recall that SDC methods aim to formalize privacy as a property of a particular database $D \in \mathcal{D}$, whereas DP methods aim to formalize privacy as a property of a particular release process on a database schema $\mathcal{D}$. This change captured an important shortcoming of SDC methods. The way they are implemented could not be disclosed transparently without revealing probabilistic information about the individuals whose data was altered due to disclosure limitation. Still, many optimal DP mechanisms rely on privacy-preserving errors in ways that depend heavily on the confidential data (e.g., \cite{reimherr2019kng,asi2020}), making $S(D) \mid D$ difficult to characterize in practice.

Should the distributions of randomized errors due to privacy depend on the confidential data? Even though the form of the mechanism can be transparently disclosed, the usefulness of this disclosure varies substantively for different mechanisms, which we explored in Section \ref{sec:sdp_count_mechs}. From these examples, the discrete Laplace mechanism provides independent perturbations to collections of statistics; because the perturbation forms a location family with independent noise set by the PLB, the distance between the private statistic $S$ and the non-private statistic $T(X)$ is independent of $X$. Aside from this relatively simple class of mechanisms, this property is not generally shared. Some primitive mechanisms, such as the exponential mechanism and its variants, do not easily allow for characterizations of errors due to privacy independent of the data. This is not to say we shoul not use mechanisms like these! It only means we should not ignore the tractability of valid downstream inference as a design consideration.

In particular, the ubiquitous use of post-processing in DP methodology yields many different methods which meet certain optimality criteria, but for which the distribution of $S(D) \mid D$ is highly data dependent and sometimes computationally intractable. This is the case for the U.S. Census TopDown algorithm, which sequentially post-processes dependent count queries to conform to global public information and various internal self-consistency rules \citep{Abowd2019}. This suggests that both theoretically \citep{seeman2022formal} and empirically \citep{Seeman2020}, DP results should be released with and without post-processing applied whenever possible.

\subsection{Challenges in Schema Choice and Data Generation}

SDP guarantees, regardless of whether using SDC or DP, depend heavily on the schema, $\mathcal{D}$. While SDP focuses on the form of the statistics we want to release, $\mathcal{S}$, the choice of $\mathcal{D}$ limits the possible values of $\mathcal{S}$ a priori. Moreover, from a system-level perspective, we tend to view $\mathcal{D}$ as a static entity, when in reality, schemas are dynamic and change over time. Schemas can grow to account for new unit attributes; e.g., many databases containing protected health information (PHI) are now updated to include information on COVID-19, such as vaccination status and testing history. Additionally, individual contributions to a database change over time, such as with streaming user data; this is an important consideration for databases regularly updated with event data, such as application logs from user behavior within different software applications. While there is some emerging work on this topic, we feel that neither SDC nor DP has developed robust solutions to these problems yet. Hence, we see this a budding area for future research opportunities.


Furthermore, SDP frameworks tend to view collected data as complete, full-information data, but rarely is this true in practice. Any social science data collection scheme could suffer from one of the many sources of ``total survey error" \citep{Groves2011} such as measurement error due to social desirability bias, errors due to missingness or other systematic non-response, or sampling procedures used to construct the database. We included these at the top of Figure \ref{fig:sdp_flowchart}, as most SDP analyses deal with human-level data. Because all information in statistical data privacy is typically taken "at face value," the practical effects of accounting for ambiguity in this process are often lost. 

Model-based SDC methods, like those discussed in Section \ref{sdc:risk}, can account for some aspects of the data generating process, like survey sampling. However, incorporating similar ideas into DP is conceptually challenging, as the methodological details themselves also depend on the confidential data \citep{bun2020controlling,seeman2021posterior}. Resolving these differences is especially important for the needs of data curators at official statistical agencies like the U.S. Census.

\subsection{Privacy and other ethical dimensions of data sciences}

Even though we have focused on data privacy in a narrow, technical sense, privacy is a naturally interdisciplinary topic which involves philosophical, legal, and political scholarly traditions. The legal operationalization of SDP remains an open problem, as there is much debate as to how SDP approaches capture different legal statutes. \cite{rogaway2015moral} argues against any approach that a priori privileges one conception of privacy over another, as all SDP methods are inherently political in the way they allocate access to different data in different forms. If we argue one kind of political allocation is automatically "better" than another, we risk ignoring how defining the terms of that allocation influences our comparisons. Science and Technology Studies (STS) scholars refer to these as ``abstraction traps," which have been studied in algorithmic fairness \citep{selbst2019fairness}. 

Additionally, SDP is but one of many research areas which attempts to imbue data analysis processes with sociologically desirable properties, such as interpretability \citep{carvalho2019machine} or fairness \citep{mitchell2021algorithmic}. Current research has pointed to limits in the ability to jointly satisfy DP guarantees and certain definitions of algorithmic fairness, both quantitatively \citep{cummings2019compatibility} and qualitatively \citep{green2021escaping}.

\subsection{Closing the Gap Between Theory and Practice}
\label{sec:disc_future}

Here, we propose directions for open research that aims to resolve ideological tensions within SDC and DP research communities and direct future research towards addressing the needs of data subjects, curators, and users simultaneously.





SDP research, in its current state, is largely focused on establishing theoretical asymptotic results. Such results are clearly valuable, as they bound the sample complexity of SDP problems in collecting and releasing private statistics. However, when practitioners are deciding which method to use, we argue that such approaches fail to support them.

Furthermore, it creates barriers to entry for who is reasonably prepared to use the results from SDP methodology. Private companies that collect data at scale and use DP, like Google and Microsoft, have enough data to estimate the regimes in which asymptotic results offer useful characterizations of privacy risk and utility. But for small datasets, like many of those in the social and behavioral sciences, such techniques are infeasible. As a research community, we ought to enable everyone to use SDP, regardless of database size.

None of our critiques should detract from the theoretical value of this work, as it is an important step towards applicability! Instead, here we highlight open questions within SDP research that take practitioner's barriers to using SDP seriously. These research directions require substantive efforts from the SDP community (computer scientists, statisticians \& data users) to help close the widening gap between SDP theory and practice, most importantly along a few key dimensions:

\begin{enumerate}
    \item \textbf{Finite-sample utility guarantees:} the close theoretical intersections between learning theory and privacy theory have motivated the sample complexity approach to SDP problems. As discussed above, this does not help practitioners easily identify the asymptotic regime in which these results apply. Future research ought to highlight tools that allow researchers working without data at scale to select optimal mechanisms for their use case. 
    \item \textbf{Uncertainty quantification:} uncertainty quantification is the foundation of statistical reasoning, and future SDP work needs to prioritize valid uncertainty quantification. For the design approach, this requires considering optimal inference in terms of total uncertainty, and not just uncertainty about the non-private estimator. For the adjustment approach, this requires examining how SDP methods influence the bias and variability of statistics produced from sanitized results. sls
    
    \item \textbf{Optimization against operational intangibles:} many of the logistical requirements for using SDP in practice require constraints that prevent optimal mechanisms from being used. For example, the U.S. Census's requirements for releasing self-consistent microdata poses problems not only for optimality, but for consistent data utility across queries \citep{abowd2021uncertainty}. Future research should treat seriously these operational requirements, like the need for microdata or interpretable error distributions. 
    \item \textbf{Computational barriers:} the focus of the majority of DP mechanisms is in optimizing the trade-off between privacy and utility. However, computational issues are usually a third, neglected dimension of the problem, as mechanisms that are optimal from a privacy-utility perspective may be computationally prohibitive to implement. These problems arise deterministically with finite computing problems \citep{mironov2012significance} as well as computing with randomized algorithms \citep{ganesh2020faster,seeman2021exact}. For example, instance-optimal mechanisms like the inverse sensitivity and $k$-norm gradient mechanisms require sampling from an intractable distribution, and failure to draw an exact sample using finite computing or finite MCMC approximation consumes additional PLB. Future research should explore trade-offs from this three-dimensional perspective, instead of the two-dimensional perspective offered by privacy vs. utility alone.
    \item \textbf{Extended trust models}: right now, the majority of trust models in SDP focus on the central model. However, techniques from secure multiparty computation (SMC) could be used to extend SDP methodology to offer more practical flexibility in trust modeling \citep{karr2010secure}. There are many possible opportunites to synthesize studies of privacy-preserving secure multiparty computation and federated learning (e.g., \cite{lindell2009secure,snoke2018providing,kairouz2021advances}).
\end{enumerate}

\section{Conclusion}

In conclusion, this review paper highlights and demonstrates the common methodological foundation of SDC and DP, and their associated quantitative and qualitative trade-offs required to investigate data privacy from either perspective. By focusing on the statistical viewpoint, SDP will produce and support the data sharing necessary for reproducible scientific discourse and democratic data governance. Whether using SDC or DP, or whether by design or adjustment, we all ought to remember that ``different roads sometimes lead to the same castle" \citep{martin2011game}.

\begin{summary}[SUMMARY POINTS]
\begin{enumerate}
\item SDC and DP methods are built upon common statistical foundations that make different but necessary compromises in conceptualizing privacy as properties of a particular database or as a schema.
\item SDP is inseparable from the study of data generating processes, as mechanism implementations introduce new privacy-preserving errors to be treated holistically alongside other error sources.
\item Both SDC and DP can suffer from model misspecification, and addressing this misspecification statistically can help improve our understanding of privacy and utility guarantees.
\item Future SDP research should address open statistical problems typically left unarticulated by theoretical SDP research, such as valid statistical inference, computational tractability, and compatibility with probability models, and their interplay. 
\end{enumerate}
\end{summary}

\section*{DISCLOSURE STATEMENT}
The authors are not aware of any affiliations, memberships, funding, or financial holdings that might be perceived as affecting the objectivity of this review. 

\section*{ACKNOWLEDGMENTS}
A.S. would like to acknowledge Vishesh Karwa and late Steve Fienberg for many early discussions and sharing of ideas on validity of privacy-preserving statistical inference; this paper is in honor to them and to our joint work that was never completed. The authors are supported in part by NSF Awards No. SES-1853209 and NCSES-BAA 49100421C0022 to The Pennsylvania State University. 

\bibliographystyle{ar-style1}
\bibliography{references.bib}

\end{document}